\documentclass[preprint]{elsart}

\usepackage{graphicx}              %%%% to include graphics 

\begin{document}

\begin{frontmatter}

\title{ Heavy-Residue Isoscaling  as a Probe of the Process of N/Z  Equilibration }

\author[cyctamu]{G.A. Souliotis},  \thanks{\footnotesize E-mail address: soulioti@comp.tamu.edu }
\author[cyctamu,iopsas]{M. Veselsky}$^{}$,
\author[cyctamu]{D. V. Shetty},
\author[cyctamu]{S. J. Yennello}.

%%************************************************************************************************
\address[cyctamu]{Cyclotron Institute,
           Texas A\&M University, College Station, TX 77843, USA }
%%************************************************************************************************
\address[iopsas]{Institute of Physics, Slovak Academy of Sciences, 
           Dubravska 9, 84228 Bratislava, Slovakia }
%%************************************************************************************************

\begin{abstract}

The isotopic and isobaric scaling behavior of the  yield ratios of heavy projectile
residues from the collisions of 25 MeV/nucleon $^{86}$Kr projectiles on  $^{124}$Sn
and  $^{112}$Sn targets is investigated and shown to provide information on the 
process of N/Z equilibration occurring between the projectile and the target. 
The  logarithmic slopes $\alpha$ and $\beta^{'}$ of the  residue yield ratios
with respect to residue neutron  number N and neutron excess  N--Z are obtained as a function
of the atomic number Z and mass number A, respectively,  whereas  excitation energies
are deduced from velocities. The relation of the isoscaling parameters  $\alpha$ and $\beta^{'}$ 
with  the N/Z of the primary (excited) projectile fragments is employed  to gain  access to the
degree of N/Z equilibration prior to fragmentation as a function of  excitation energy.
%%%%
A monotonic relation between the N/Z difference of fragmenting quasiprojectiles
and their excitation energy is obtained indicating that  N/Z equilibrium is approached 
at the highest  observed excitation energies. %%%%  E$^{*}$/A $\sim$ 2.2 MeV/nucleon. 
%%%%
Simulations  with a deep-inelastic transfer model  are in overall  agreement with the isoscaling
conclusions.  %%%%%%%%%% up to excitation energy of $\sim$ 2.0 MeV/nucleon. 
%%%% whereas beyond this the DIT  predicts that N/Z equilibration is attained  at larger excitation
%%%% energies, thus longer  interaction times.
The present residue isoscaling  approach to N/Z equilibration  offers  an attractive  
tool of  isospin and reaction  dynamics studies in collisions
involving beams of  stable or rare isotopes.   

\end{abstract}

\begin{keyword} Heavy Residues, N/Z equilibration, Deep-inelastic transfer, Isoscaling, 
                Fermi Energy, Nuclear reactions

\PACS 25.70.Mn,25.70.Lm,25.70.Pq

\end{keyword}

\end{frontmatter}

%**************************************************************************
\section{Introduction}
%**************************************************************************

The recent  progress in the field of Rare Isotope Beams (RIB) worldwide %%%\cite{}
opens up new and exciting frontiers in the studies of nuclear structure and nuclear reactions
(e.g. \cite{LongRP,EURISOL}).
%%%
For the latter in particular, the availability of beams with large neutron-to-proton ratio,
N/Z, provides the   opportunity to explore the collision dynamics of  very 
isospin-asymmetric nuclear systems \cite{Bao1}. 
In such reactions, the N/Z  degree of freedom (loosely called ``isospin'' degree of freedom)
and its equilibration  have prominent
roles and can serve as valuable probes of the symmetry energy term of the nuclear 
equation of state \cite{Bao2,Bao3}.
Extensive work in deep-inelastic heavy-ion collisions at low energies (E/A$<$10 MeV, 
for a review see, e.g., \cite{Volkov,Udo}) provided insight into the relevant reaction mechanisms 
\cite{Fre,Petro,Planeta}. 
At these energies, the collisions are of binary character involving 
a transient dinuclear complex. 
%%%%%%%%%%%%
A fast redistribution of neutrons and protons takes  place between the interacting projectile
and target %%%% during the interaction time; 
and characterizes the process of N/Z (or isospin) equilibration (or relaxation), governed predominantly
by the potential energy surface of the dinuclear complex.
At medium energies, recent  work \cite{SJY1,SJY2} indicated that essentially above the 
Fermi energy  isospin equilibrium  is not reached even for the most damped (central) collisions.
In these studies,  isotopic \cite{SJY1} or isobaric \cite{SJY2} yield ratios of selected 
intermediate-mass fragments (IMF) were used to probe  isospin equilibration in
near-central collisions.
Insight into isospin equilibration has also been provided in a recent study
of reconstructed quasiprojectiles undergoing  multifragmentation
following a deep-inelastic collision \cite{MV_SiSn}.
%%%%
In addition, IMF isotopic yield ratios  were employed in a recent work to probe 
the degree of N/Z equilibrium in peripheral collisions of $^{124,112}$Sn+$^{124,112}$Sn
at 50 MeV/nucleon \cite{Tsang_diffusion}. Parallel studies have also been performed at relativistic
energies \cite{Rami}.  A recent theoretical treatment of N/Z transport and equilibration is
given in \cite{Pawel}.

It would be of great interest  to probe experimentally the evolution towards N/Z equilibrium as a
function of  the excitation energy (or the impact parameter) from the very peripheral to the most 
damped collisions. In such an endeavor, the properties of heavy residues appear  be of central 
importance.  As well known, in deep-inelastic collisions at low and medium energies, a large fraction
of the reaction cross section corresponds  to the production of heavy residues, i.e.,
large surviving remnants of the projectile (or target) after  evaporation of neutrons, protons or
 light charged particles  \cite{Fuchs}.
The velocities of the  residues, strongly correlated to their mass, provide information on the excitation
energy imparted  into the collision partners. %%%\cite{}.
The N/Z of the residue is  reminiscent of  the N/Z of the corresponding hot primary
fragment, but the former is substantially  affected by the evaporation process.
Efforts  to reconstruct the primary quasiprojectile \cite{Petro,Planeta} based on evaporation
codes provide a first picture of the process of  N/Z equilibration as a function of
the excitation energy  and thus, the interaction time.
It should be noted that information on N/Z equilibration  has also been obtained 
recently  in prompt  $\gamma$-ray  studies \cite{Pierrou} in which enhanced emission
of dipole $\gamma$ rays during the N/Z equilibration process was  observed in isospin 
asymmetric reactions.

In the present study, we will describe a new  experimental approach to follow the process
of N/Z  equilibration  based on  heavy-residue isoscaling.  The approach employs yield
ratios of  isotopically resolved  heavy residues obtained  from two  isospin-asymmetric
reactions in conjunction with high-resolution measurement of residue velocities.
This Letter is organized as follows. In section 2,  a concise description of the experimental  measurements
of heavy projectile residues  from   the reactions of  $^{86}$Kr (25MeV/nucleon) 
with $^{124}$Sn and $^{112}$Sn targets is given. In section 3,  the analysis of residue yield ratios
and  the  procedure to  extract information about  N/Z equilibration are described,
followed by model comparisons. 
Finally a summary and  conclusions are given in Section 4.

%%%************************************************************************
%**************************************************************************
 \section{Experimental Details}
%**************************************************************************

The experimental measurements and analysis procedures are  described in detail
in our  recent article \cite{GS_isoscale} in which  a detailed   study of isotopic
scaling  of heavy projectile fragments from 25 MeV/nucleon  $^{86}$Kr-induced
reactions on $^{124,112}$Sn and $^{64,58}$Ni is reported.
Herein we present a subsequent  analysis of the   Kr+Sn  data with the aim of  obtaining
detailed information on the N/Z equilibration process. 
%%%
%%%-------------------------------------------------------------------------------------------
%%% As indicated in  \cite{GS_isoscale}, the use of the  high-resolution recoil separator  MARS
%%% in combination with  $\Delta$E--E and time-of-flight techniques
%%% provided  information  on the atomic number Z, the ionic charge q, the mass number A
%%% and the velocity distributions of the projectile fragments.
%%%-------------------------------------------------------------------------------------------
%%%
For completeness, a short description of the  experimental conditions follows.
The measurements were performed at the Cyclotron Institute of Texas A\&M
University.
%%% following the general experimental scheme described in our recent 
%%% articles \cite{GS1,GS2,GS3}.
A  25 \hbox{MeV/nucleon} $^{86}$Kr$^{22+}$ beam  ($\sim$1 pnA) from 
the K500 superconducting cyclotron  interacted with  $^{124}$Sn and  $^{112}$Sn
targets. Projectile residues were analyzed with  the MARS recoil separator \cite{MARS}.
The primary beam struck the target at an angle of  4.0$^{o}$ relative to the optical 
axis of MARS.
%%The angular acceptance of MARS  allowed 
%%projectile fragments to be  accepted inside   the grazing angle  of these  reactions.
At the focal plane of MARS, the fragments were collected in a 5$\times$5 cm  
$\Delta $E--E  Si detector telescope. 
Time-of-flight was measured between two PPACs (parallel plate avalanche 
counters) placed at  the MARS dispersive image and at the focal plane,
respectively.
The horizontal position provided by the first PPAC  and  the field measurement of the 
MARS first dipole magnet  were  used to determine the magnetic rigidity,  
$B\rho$. 
The reaction products were  characterized event-by-event using 
energy-loss, residual energy, time-of-flight, and $B\rho$. 
These quantities were calibrated with  a low intensity $^{86}$Kr beam
and other beams at 25 MeV/nucleon. 
With the procedures described in Ref. \cite{GS_isoscale,GS_PLB}, the atomic number Z, 
the ionic charge q, the mass number A  and the velocity  of the fragments 
were obtained with high resolution.
The resolutions of Z, q and A were 0.5, 0.4 and  0.6 units, respectively, for near
projectile residues.
Summation over the ionic charge states  provided yield distributions with respect
to  Z, A and  velocity from which the yield distributions employed in this work 
were obtained.
We note  that the measurements were performed in the magnetic rigidity
range of 1.3--2.0 T\,m (by superposition of successive settings of the separator)
and angular range of 2.7$^o$--5.4$^o$ \cite{GS_isoscale}  which lies
inside the grazing angle of 6.5$^o$ of the Kr+Sn systems at 25 MeV/nucleon.
This B$\rho$ and angular range enabled efficient collection of heavy projectile
residues produced in a large  range of energy damping, from quasielastic to 
deep-inelastic collisions.  

%**************************************************************************

\section{Results and Discussion}

%**************************************************************************

Before discussing  residue isoscaling and N/Z equilibration,
we will  examine the characteristics of  residue velocities and the 
information on excitation energy that they can provide.
It is well known  that peripheral reactions between massive nuclei around the Fermi energy
\cite{Fuchs} proceed via a deep-inelastic transfer mechanism involving substantial nucleon
exchange \cite{MV_SiSn,HR1,HR2}.
This mechanism is responsible for the creation of highly excited primary products  that 
de-excite to produce the observed fragments.
Information on  the excitation energy of the primary fragments  from the present reactions
can be obtain from  the measured velocity versus mass number A  (or atomic number Z) correlations. 
Fig. 1a  presents the average
velocities of the projectile fragments as a function of  A. 
Closed symbols correspond to the reactions 
with the neutron-rich $^{124}$Sn  target and open symbols to those with the neutron-poor
$^{112}$Sn target. (As we observe, the average velocities from the reactions with the two
targets are, within the experimental uncertainties, roughly  the same.)
In this figure, we observe that for fragments  close to the projectile, the velocities
are slightly below that of the projectile, corresponding to very peripheral, low-excitation energy
events. A monotonic decrease of velocity with decreasing A  is observed,
indicative of larger dissipation and thus, higher excitation energies.
A similar correlation  of velocity with the atomic number Z
of the fragments is obtained, as presented in Fig. 1a of \cite{GS_isoscale}.

The  descending velocity--mass  correlation continues  down to A$\sim$65.
For lower masses,  the average velocity appears to increase slowly.
As pointed out in \cite{GS_isoscale},
the observation  of a  minimum  velocity for A$<$65 (Z$<$28) indicates that  
these residues originate from primary fragments with a maximum observed  
excitation energy. Fragments with A near the projectile down to A$\sim$65  originate from 
evaporative type of de-excitation  which preserves, on average, the emission direction
of the residues. In this case, the residue velocity can be employed to obtain  excitation energy.
%%%
Residues with lower A may arise from cluster emission or multifragmentation   and the
velocity of the inclusively measured fragments is not  monotonically  correlated with  excitation
energy and mass \cite{GS_isoscale}. (These lower mass fragments are not the focus
of the present work and are omitted from Fig. 1).
%%
%%--------------------------------------------------------------------------------------------------
%% The ascending  part of the velocity vs mass  correlation for the lower part of the mass range is
%% primarily due to the combined effect of angle and  magnetic rigidity selection \cite{MV_SnAl}.
%% The forward angle range (2.7$^o$--5.4$^o$)  results in the selection of either the forward or the
%% backward kinematical solution in the moving frame of the quasiprojectile undergoing 
%% cluster-emission or multifragmentation, whereas the magnetic rigidity range (1.3--2.0 T\,m)
%% results in the selection of the forward solution.
%%--------------------------------------------------------------------------------------------------
%%

Employing the observed average residue  velocities  for the  Kr+Sn syetems 
and, furthermore, assuming binary   kinematics and equal division of excitation energy
(which is a reasonable approximation for the present reactions
\cite{MV_SiSn,Madani}), 
we can estimate an  average excitation energy per nucleon for the hot quasiprojectile 
fragments as a function of mass  as presented in Fig. 1b.
Using the Fermi gas relationship $ E^* = \frac{A}{K} T^2 $, 
with  T  the tempretature and K the inverse level density parameter,  taken as K=13 MeV \cite{JBN1},
we can also estimate the  temperature of the Kr-like quasiprojectiles.
We note that,  at the maximum observed excitation energy of 2.2 MeV/nucleon, 
the temperature is 5.3 MeV, which is near  the threshold for multifragmentation
\cite{MV_SiSn,JBN1,MV_SnAl}.

%%%**************************************************************************************************
%%%**************************************************************************************************
%%% Isoscaling:

Having  presented the excitation energy characteristics  of the measured residue data,
we will examine  the isoscaling properties of their  yields.
It has been shown  \cite{Tsang1,Tsang2,Tsang3,Botvina}, that the ratio R$_{21} = $ $Y_{2}(N,Z)/Y_{1}(N,Z)$
of the yields of a given fragment (N,Z) from two reactions with similar excitation energies
and similar masses, which differ only  in N/Z,  eliminates the effects of secondary decay
and provides information about the excited primary fragments.
An exponential relation with respect to N and Z of the form:  
%%%%%%%%%%%%%%%%
\begin{equation}
     R_{21}(N,Z) = Y_{2}(N,Z)/Y_{1}(N,Z) = C \exp(\alpha N  + \beta Z)
\end {equation}     
%%%%%%%%%%%%%%%%  
has been  obtained both experimentally and theoretically  and has been  termed isotopic scaling
or isoscaling.
In the framework of the grand canonical ensemble, the scaling parameters 
$\alpha$ and $\beta$ are expressed as
$\alpha$ = $\Delta \mu_{n}$/T and $\beta$ = $\Delta \mu_{p}$/T, with 
$\Delta \mu_{n}$ and $\Delta \mu_{p}$ being the differences in the neutron  and the 
proton chemical potentials and T the temperature  of the fragmenting systems \cite{Tsang1}.
C is an  overall normalization constant.

%%%%-----------------------------------------------------------------------------------------------------
%%%% ISOTOPIC Scaling here: 

For the Kr+Sn  data, we construct the  yield ratio $R_{21}(N,Z)$ using the
usual convention that index 2 refers to the more neutron-rich system and index 1 
to the less neutron-rich one.
Fig. 2a  shows the isotopic yield ratios  R$_{21}$(N,Z) as a function of fragment
neutron number N for several  isotopes. As also described in \cite{GS_isoscale}, for each Z,
exponential functions  of the form $ C exp(\alpha N) $ were  fitted to the data and are shown
in Fig. 2a  for the selected isotopes. 
In  Fig. 2b,  we present  the slope parameter $\alpha$  of the
exponential  fits  as a function of Z.
As already pointed out in \cite{GS_isoscale}, the parameter $\alpha$ remains roughly
constant at an average  value of 0.43$\pm$0.01 for elements up to Z$\sim$26  
(corresponding  to primary events with the maximum observed  excitation energy of 2.2 MeV/nucleon)
and  decreases gradually for larger fragments.

%%%%-----------------------------------------------------------------------------------------------------
%%%% ISOBARIC Scaling here: 

In addition to the  isotopic scaling discussed above,
we will  present the scaling relation  in an alternative expression,
namely,  as isobaric scaling.
As pointed out  by Botvina \cite{Botvina}, the isoscaling relation of Eq. 1  can also be expressed 
as: 
%%%%%%%%%%%%%%%%
\begin{equation}
    R_{21}(A,N-Z) = C \exp \{ \alpha^{'} A  + \beta^{'} (N - Z) \}
\end {equation}       
%%%%%%%%%%%%%%%%
with $\alpha^{'} = ( \alpha + \beta )/ 2 $ 
and  $\beta^{'}  = ( \alpha - \beta )/ 2 $.
This  isobaric scaling relation expresses   an exponential dependence
of the yield ratios of a given fragment of  an  isobaric chain A
on the neutron excess   N--Z  (or, equivalently, on the third component of the isospin 
t$_z$ = (N--Z)/2).  
%%%%%%%
In a manner similar to isotopic scaling, the scaling parameters 
$\alpha^{'}$ and  $\beta^{'}$ express differences of chemical potentials
$\mu_{A}$ and $\mu_{N-Z}$ conjugate to the variables A and N--Z.
%%%%%%%% that characterize the two-component nuclear system.
These potentials are connected to $\mu_{n}$ and $\mu_{p}$ via
the relations: $ \mu_{A}   = (\mu_{n}+\mu_{p})/2 $  and
               $ \mu_{N-Z} = (\mu_{n}-\mu_{p})/2 $. These expressions
lead to the following relations for        %%%%%%%%%%% $\alpha^{'}$ and  $\beta^{'}$:
the isobaric scaling parameters:
               $ \alpha^{'} = \frac{\Delta (\mu_{n}+\mu_{p})}{2T} $ and
               $ \beta^{'}  = \frac{\Delta (\mu_{n}-\mu_{p})}{2T} $.

In the isobaric scaling expression of Eq. 2, the dependence on A via $\alpha^{'}$ is weak,
since the   coefficient  $\alpha^{'}$  is close to zero \cite{Botvina} and it will not be
considered in the present study.
From the  data, we construct the yield ratio $R_{21}(A,N-Z)$ again with  the 
convention that  index 2 refers to the more  neutron-rich system.
Fig. 3a shows the yield ratios  R$_{21}$(A,N--Z) as a function of fragment neutron-excess 
N--Z for several isobars. In this figure, we see  that  the ratios for each isobaric  chain
exhibit a remarkable  exponential behavior. For each A, an exponential function 
%%%%%%%%of the form  $ C exp( \alpha (N-Z)) $
was fitted to the data and also shown in Fig. 3a  for the selected masses. 
%******************************************************************************
%
In  Fig. 3b,  we present  the slope parameter $\beta^{'}$  of the
exponential  fits  as a function of A.
The slope parameter $\beta^{'}$ remains roughly  constant at an average  value of
0.47$\pm$0.01  for A $<$60   and decreases monotonically for A$>$ 60.
%%%As discussed earlier, the fragment range A=25--60 corresponds to primary events
%%%with the maximum observed  excitation energy of 2.2 MeV/nucleon \cite{GS_isoscale}.
We note  that the isobaric scaling parameter $\beta^{'}$  herein obtained
for the low-mass range is in very good agreement with the expected value
of $ ( \alpha - \beta )/ 2 $ if the experimentally obtained values of the 
isoscaling parameters  $\alpha$=0.43$\pm$0.01 and  $\beta$=--0.50$\pm$0.01 
 \cite{GS_isoscale} are used.

In the following, we will use the relation of the isoscaling parameters
$\alpha$ and  $\beta^{'}$ with the primary fragment N/Z to extract information
on N/Z equilibration.
In the framework  of the grand canonical approximation
of the  statistical multifragmentation model (SMM) \cite{Botvina},
the isoscaling parameters  $\alpha$ and  $\beta^{'}$  can be directly 
related to the coefficient
C$_{sym}$ of the  symmetry energy term of the nuclear binding energy
and the N/Z of the primary fragmenting systems. 
The following expressions are  obtained:
%%%
\begin{equation}
     \alpha = 4 \, \frac {C_{sym}}{ T } ( (\frac{Z_1}{A_1})^2  -  (\frac{Z_2}{A_2})^2    )
\label{a1}
\end {equation}       
%%%
%%%and,
%%%
\begin{equation}
     \beta^{'} = 4 \, \frac {C_{sym}}{ T } ( \frac{Z_1}{A_1}  -  \frac{Z_2}{A_2}    )
\label{Csym1}
\end{equation}       
%%%
in which $Z_1$,$A_1$ and  $Z_2$,$A_2$ refer to the fragmenting quasiprojectiles from reactions 1 and 2,
respectively.
%%%
From these equations, after some manipulation we obtain:
%%%
\begin{equation}
     \alpha = 8 \, \frac{C_{sym}}{ T }  (\frac{Z}{A})_{ave}^3  \Delta( \frac{N}{Z} )_{qp}
\label{Csym}
\end {equation}       
%%%
%%%and
%%%
\begin{equation}
     \beta^{'} = 4 \, \frac{C_{sym}}{ T }  (\frac{Z}{A})_{ave}^2  \Delta( \frac{N}{Z} )_{qp}
\label{Csym2}
\end {equation}       
%%%
where $(Z/A)_{ave}$ is the average Z/A of the quasiprojectiles, taken to be the average
Z/A of the composite systems $^{86}$Kr+$^{124}$Sn and $^{86}$Kr+$^{112}$Sn,  and
$ \Delta( N/Z )_{qp} $ expresses the N/Z difference of fragmenting quasiprojectiles corresponding 
to a given value of fragment Z or A, scaling parameters  $\alpha$ or $\beta^{'}$ and the corresponding 
temperatures T.
%%%
Assuming that fragmentation occurs at  normal density, using  C$_{sym}$ = 25 MeV 
\cite{Botvina},  the  $\alpha$ and $\beta^{'}$ values obtained from the isotopic (Fig. 2b) and
isobaric (Fig. 3b) scaling fits, respectively,  and temperatures determined from excitation energies,
we can determine the values of  $ \Delta( N/Z )_{qp} $ as a function of the observed fragment 
atomic number Z, as well as,  mass number  A.
Subsequently, plotting   $ \Delta( N/Z )_{qp} $ versus the average E$^{*}$/A value corresponding to 
each Z  or  A, we obtain the correlations  presented in Fig. 4a. 
The open circles correspond to the isotopic scaling procedure, whereas the full circles to the 
isobaric scaling procedure.
In this figure, the horizontal dotted line expresses the N/Z difference of
fragmenting quasiprojectiles under the condition of  isospin equilibrium.

The points from the isotopic scaling approach
are in very good overall agreement with the  values obtained from the  isobaric scaling 
approach, in practice indicating  the equivalence of these two procedures. 
It should be noted that in the latter procedure,
because of the larger (more than double) number of integers A, as compared to Z,
a more detailed mapping of the primary  N/Z  vs  E$^*$ correlation is achieved
in  the excitation energy region from very peripheral to  more  damped collisions
(Fig. 4a).

%------------------------------------------------------------------------------------------------------
%% Comments, explanations etc:

The correlations  presented in Fig. 4a show  the evolution of the N/Z equilibration process 
in the present isospin-asymmetric collisions. The monotonic increase of 
$ \Delta( N/Z )_{qp} $ with excitation energy can be understood as a result of the 
mechanism of nucleon exchange.
Fragments close to the projectile are  produced in very peripheral collisions
in which a small number of nucleons is exchanged and thus, the N/Z difference of the fragmenting 
quasiprojectiles from  $^{86}$Kr+$^{124}$Sn and $^{86}$Kr+$^{112}$Sn is small. Fragments progressively
further from the projectile originate from  collisions with larger projectile--target overlap in which 
a large number of nucleons is exchanged and their excitation energy is higher.
In such cases,   the N/Z difference of the fragmenting  quasiprojectiles is progressively larger,
eventually approaching, for this  energy regime,  the N/Z difference corresponding to isospin
equilibration. The conclusion that  N/Z equilibration is attained
for the most damped collisions is rather expected in accord with recent  BUU calulations
\cite{Bao3}.
The experimental determination of the $ \Delta( N/Z )_{qp} $ vs E$^{*}$/A correlation (Fig. 4a) is
the basis of the present approach to study the process  of N/Z equilibration. As we have indicated, it is 
primarily based on the N/Z information provided by the isoscaling parameters $\alpha$ and $\beta^{'}$.
In addition, it utilizes the connection
of the average excitation energy with fragment size (Z or A)  ensued by the binary character
of the collisions.
%-------------------------------------------------------------------------------------------------------
%%% DIT model

It would be instructive to compare the results of N/Z -- E$^*$/A correlations
obtained from the present data via the isoscaling approach  with model simulations
appropriate for this energy regime.
In Fig. 4b, we show the predictions of the deep-inelastic transfer (DIT) code of Tassan-Got
\cite{DIT} which has recently been successfully applied in a variety of studies at Fermi energies
(e.g. \cite{MV_SiSn,GS_PLB,MV_SnAl}).
The DIT model simulates stochastic nucleon exchange in a Monte Carlo fashion. 
%% in the angular momentum range $\el$=250--750.
%% Events corresponding to trajectories in which the projectile--target overlap exeeded 3 fm
%% were rejected.
The predicted  average N/Z--E$^*$/A correlations  for the hot quasiprojectiles from 
$^{86}$Kr+$^{124}$Sn and  $^{86}$Kr+$^{112}$Sn are shown by the upper and lower full lines,
repectively. The upper and lower dotted lines  indicate the N/Z values expected for fully
equilibrated quasiprojectiles, whereas the horizontal full line indicates the N/Z of the 
projectile.
%%%%%%%%%%%
Using these correlations we obtain the $ \Delta( N/Z ) $ -- E$^{*}$/A correlation
shown by the full line in Fig. 4a.
%%%%%%%%%%%
As we see in this figure, the results of the isoscaling procedure seem to follow the 
model prediction up to  E$^*$/A$\sim$ 2.0 MeV, whereas beyond this energy,
the DIT model indicates  that N/Z equilibration is reached  at larger excitation energies.
%%%
We think that this  difference is mainly  due to the inability of the residue velocity  approach
to determine correctly the excitation energies near and beyond the onset of multifragmentation
(E$^{*}$/A  $\sim$ 2.2 MeV/nucleon). Kinematical reconstruction of the quasiprojectiles
is necessary to determine  these larger excitation energies (see, e.g. \cite{MV_SiSn}).
%%%%%%
Finally, we note that results similar to those of the DIT code are obtained with another  
widely used model, the nucleon exchange model (NEM) of Randrup \cite{NEM},
shown  in Figs. 4a,b by the thin dashed lines.

%-----------------------------------------------------------------------------------------------

From the aforementioned discussion, we see that  
we can follow in detail the evolution of the N/Z equilibration process
as a function of excitation energy from the quasi-elastic regime up to very  damped collisions,
provided that surviving residues are produced whose velocity is correlated to their size.
%%%
Due to the mechanism of nucleon exchange, the assumption of equal division of excitation energy 
between the projectile and the target is a good approximation for not too long 
interaction times (and, thus, not too high excitation energies). [At the longest
interaction times (highest excitation energies), the dinuclear system will  approach 
thermal equilibrium.]  The maximum observed  excitation energies inferred from residue velocities
are $\sim$2.2 MeV/nucleon,  which are moderate compared to the highest (5--6 MeV/nucleon) that
can be attained in the most damped collisions  for the present systems.
%%%
%%% In addition, the Kr+Sn systems are  not too mass  asymmetric, so equal division of excitation
%%% energy is an appropriate approximation for the present purposes.
%%%
We also point out that, as discussed in detail in the work of Botvina et al. \cite{Botvina},
grand canonical statistics at its low temperature limit can be used as an approximation to 
describe residue deexcitation  on which  Eqs. 3--6  are based.  
%%%%
Another question that might be raised from the present  treatment of data  spanning a wide
range of energy dissipation is  why  non-equilibrium in the N/Z degree of freedom at the
separation stage  is consistent with an equilibrium statistical  description
(here, in a grand canonical approximation).
%%%%
We point out  that  the degree of N/Z equilibrium at the separation stage is determined
by the dynamics  of the collision and the corresponding  interaction time.
%%%
The grand canonical statistics is applied to the de-excitation of the hot quasiprojectile residue
under the assumption (usual for treatment of  de-excitation) that enough time has elapsed after
the separation  so that its degrees of freedom are equilibrated.  
Thus, the excited  quasiprojectile is assumed to be characterized by a value of N/Z, along with its
excitation energy and spin; for this hot residue  the grand canonical statistics (in its low temperature limit)
can be  applied.

%-------------------------------------------------------------------------------------------------

It would be very interesting to study not only the evolution of the N/Z
equilibration process in a variety of isospin asymmetric systems,  but also the transition from  
N/Z equilibration to non-equilibration as a function of the projectile energy.
%%%
It may also be noted that the present  approach (with the possible inclusion of  kinematical
reconstruction of quasiprojectiles) can 
be efficiently   applied to collisions at the  limits of  N/Z asymmetry,  taking
advantage of  current and future  rare isotope beam facilities.
%%% 
Studies of the N/Z degree of freedom with the present approach can also complement
studies employing intermediate (IMF) or 
light charged particle (LCP) yield ratios from hot primary fragments undergoing multifragmentation
\cite{SJY1,SJY2,Tsang_diffusion}.
%%%
Finally, it may be pointed out that detailed studies of the N/Z degree of freedom and its 
equilibration in reactions around the Fermi energy can offer  a quantitative testing 
ground  of current transport  models of heavy-ion reactions
(e.g. \cite{IBUU,IQMD,AMD}) and provide information  on the  symmetry energy  part of the 
nuclear equation of state.

\section{Summary and Conclusions}

In summary, a new experimental  approach  to study the process of N/Z equilibration
has been  presented. The approach is based on the N/Z  information contained in the
isotopic and the  isobaric yield ratios of heavy residues from two  isospin-asymmetric
deep-inelastic collisions.
%%%
The corresponding isoscaling parameters  $\alpha$ and $\beta^{'}$ of the  residue yield ratios
with respect to residue neutron  number N and neutron excess  N--Z are obtained as a function
of the atomic number Z and mass number A, respectively. Residue excitation energies
are deduced from velocities. The relation of the isoscaling parameters  $\alpha$ and $\beta^{'}$ 
with  the N/Z of the primary (excited) projectile fragments provides  access to the
degree of N/Z equilibration prior to fragmentation as a function of  excitation energy.
Simulations  with a  deep-inelastic transfer model are in agreement with the isoscaling data,
at low and moderate excitation energies, whereas limitations in the excitation energy 
determination  near and above the multifragmentation threshold
may be responsible for an observed disagreement 
between the data and the model.
%%%
The present residue isoscaling  approach may  offer a sensitive probe  of isospin and reaction
dynamics studies in collisions involving stable,  as well as,  rare isotope beams.   

%%%%------------------------------------------------------------------------
%**************************************************************************
\section{Ackowledgements}
%**************************************************************************

\par

%%We  wish to thank the Cyclotron Institute staff for the
%%excellent beam quality. 

We wish to thank to A. Botvina  for insightful discussions.
We are also thankful to  L. Tassan-Got for using the DIT code and
J. Randrup for using the NEM code. This work was supported in part by the
Robert A.  Welch Foundation through grant No. A-1266, and the Department of Energy
through grant No. DE-FG03-93ER40773. M.V. was also supported through grant
VEGA-2/1132/21 (Slovak Scientific Grant Agency).

%%*************************************************************************
%%**************************************************************************
%%**************************************************************************
%%**************************************************************************
%%%%\bibliography{nzeq.bib}
%%%%------------------------------------------------------------------------

%%**********************************************************************************************
%%**********************************************************************************************
%%**********************************************************************************************

%Fig 1         

    \begin{figure}[h]                                        %%% htbp

    \begin{center}

   \includegraphics[width=0.60\textwidth, height=0.50\textheight ]{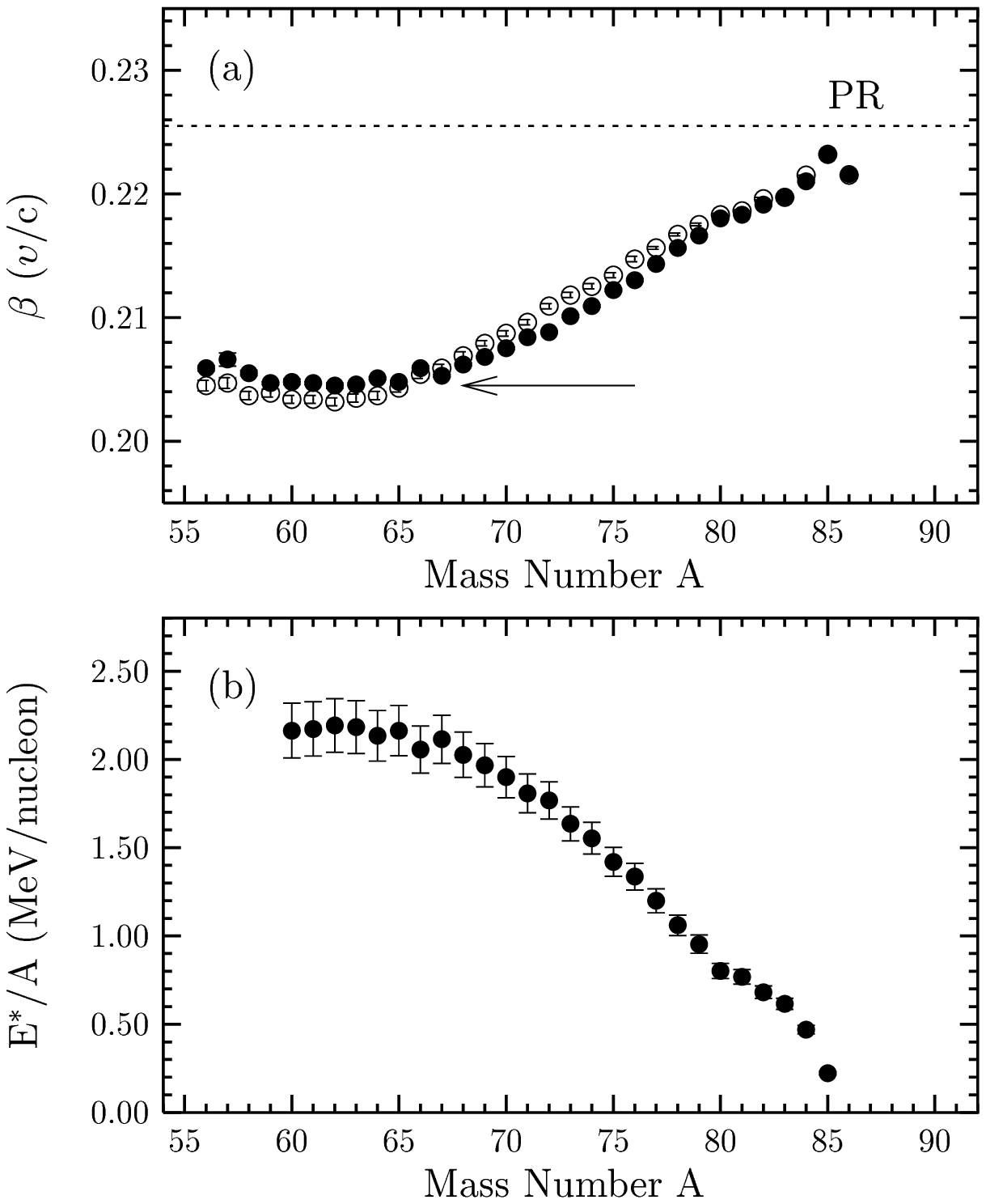}

    \caption{
           (a) Average velocity versus mass number A correlations  for projectile residues 
           from  the reactions of  $^{86}$Kr (25MeV/nucleon) with $^{124}$Sn and $^{112}$Sn.
	   Full circles represent the data with the $^{124}$Sn target 
           and open circles those with  the  $^{112}$Sn target.
           The dashed line (marked ``PR'') gives the projectile velocity, whereas
           the arrow indicates the minimum average residue velocity.
           (b) Excitation energy per nucleon evaluated from residue velocities (see text).
           }
    \label{vel}
    \end{center}

    \end{figure}

\pagebreak

%%**********************************************************************************************
%% Fig 2.

    \begin{figure}[h]                                        %%% htbp
    \begin{center}

    \includegraphics[width=0.65\textwidth, height=0.60\textheight ]{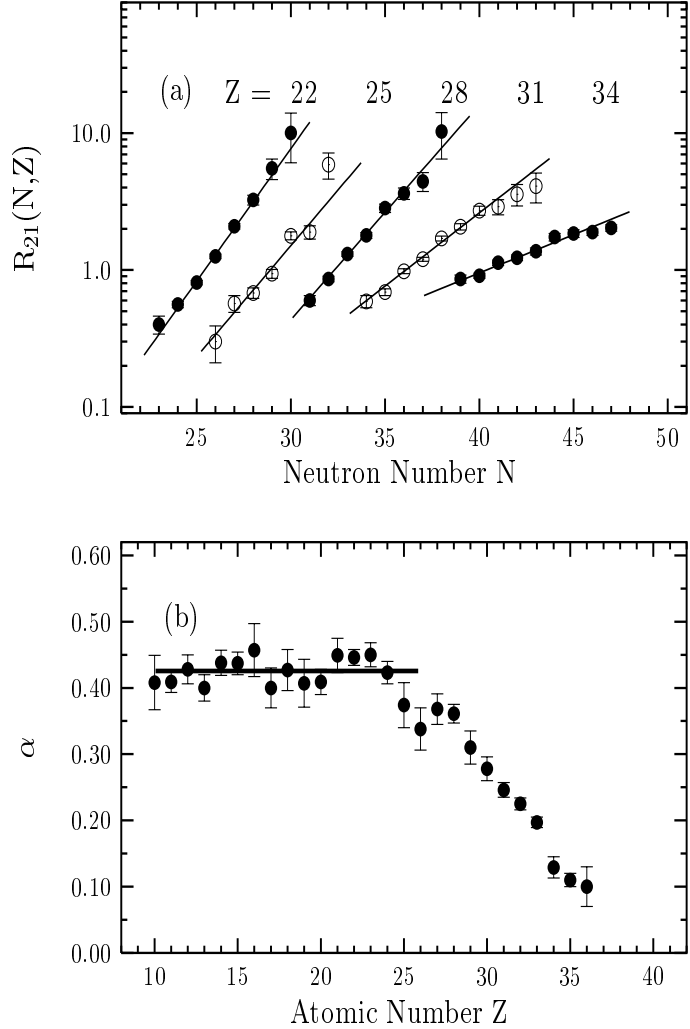}

    \caption{
           (a)Yield ratios $ R_{21}(N,Z) = Y_{2}(N,Z)/Y_{1}(N,Z) $ of  projectile residues 
           from  the reactions of $^{86}$Kr (25MeV/nucleon) with $^{124,112}$Sn  with respect
           to N for the Z's indicated. The data are given by alternating filled and open circles,
           whereas the lines are exponential fits.
           (b)
           Isotopic scaling parameter $\alpha$ as a function of Z
           (closed circles).  The straight line is a  constant value fit for the
           lighter fragments Z=10--26 (see text).
           }
    \end{center}
    \label{isoscale}
    \end{figure}

\pagebreak

%%**************************************************************************
%Fig 3        

    \begin{figure}[h]                                        %%% htbp
    \begin{center}

    \includegraphics[width=0.65\textwidth, height=0.60\textheight ]{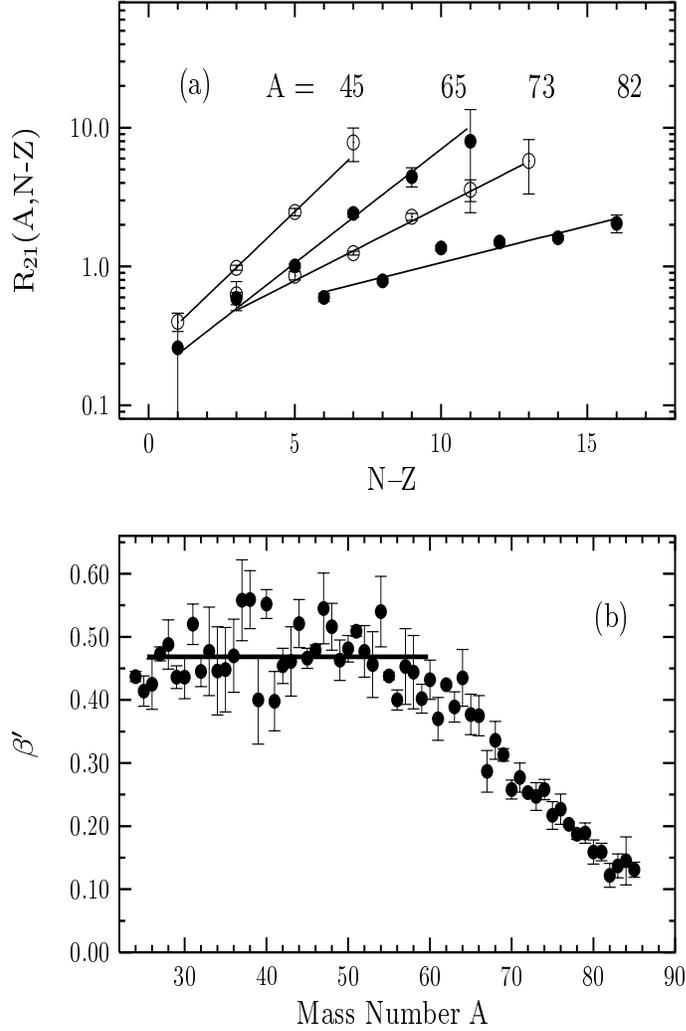}

    \caption{
           (a) Yield ratios $ R_{21}(A,N-Z) = Y_{2}(A,N-Z)/Y_{1}(A,N-Z) $ of  projectile residues 
           from  the reactions of $^{86}$Kr (25MeV/nucleon) with $^{124,112}$Sn  with respect
           to N--Z  for the masses  indicated.
           The data are given by alternating filled and open circles, whereas the lines are
           exponential fits.
           (b) Isobaric scaling parameter $\beta^{'}$ as a function of A.
           The straight line is a  constant value fit to the data
           in the lighter residue range A=25--60 (see text). }
           
    \end{center}
    \label{isoscale}
    \end{figure}

\pagebreak

%***************************************************************************************************
%%%% Fig 4

    \begin{figure}[h]
                                        %%% htbp
    \begin{center}
    \includegraphics[width=0.65\textwidth, height=0.60\textheight ]{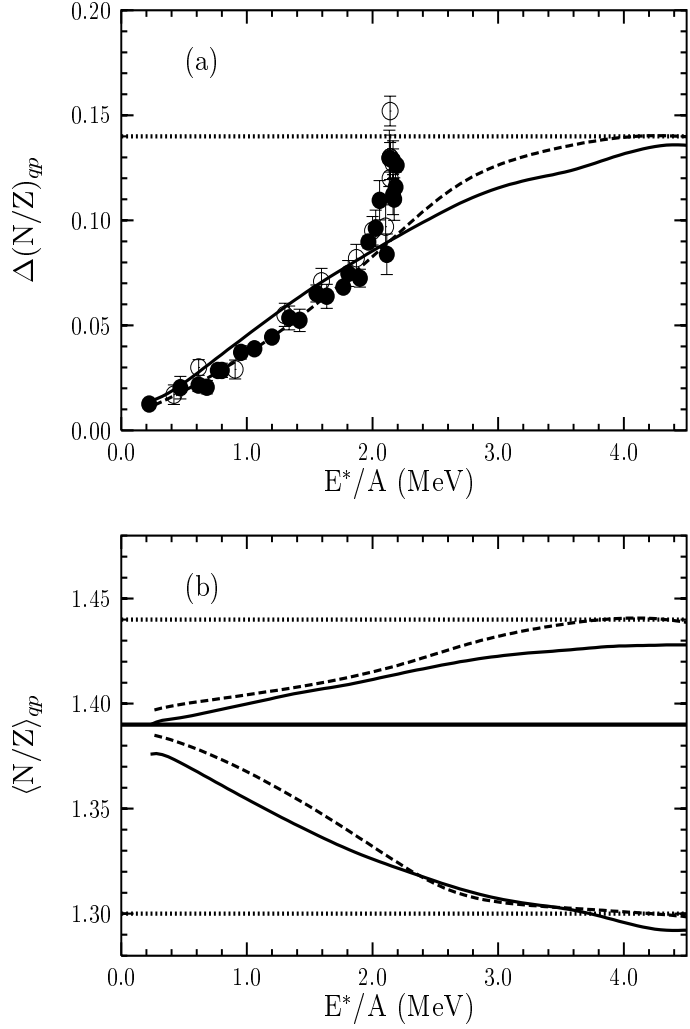}

    \caption{
               (a) Difference in N/Z of fragmenting quasiprojectiles  obtained  using Eqs. 5, 6 (see text)
                  as a function of excitation energy per nucleon  for projectile residues from  
                  the reactions  $^{86}$Kr (25MeV/nucleon)+$^{124,112}$Sn.
                  Open circles: isotopic scaling; closed circles: isobaric scaling.
                  The full and dashed lines are predictions of the DIT model \cite{DIT}
                  and the NEM model \cite{NEM}, respectively. The horizontal dotted line gives the N/Z difference
                  of isospin-equilibrated  fragmenting quasiprojectiles.
               (b) Model predictions of average N/Z of fragmenting quasiprojectiles used to get the differences
                   shown in panel (a).
                   As above: full lines:  DIT model; dashed lines: NEM model.
                   The horizontal full line indicates the projectile N/Z and 
                   the horizontal dotted  lines  give  the N/Z values of isospin-equilibrated 
                   quasiprojectiles.        
           }
    \end{center}
    \label{nzeq}
    \end{figure}

%%**************************************************************************
%%**************************************************************************
%**************************************************************************

\end{document}